# Propagation of optical excitations by dipolar interactions in metal nanoparticle chains

W. H. Weber* and G. W. Ford

Physics Department

The University of Michigan

Ann Arbor, MI 48109-1120

ABSTRACT

Dispersion relations for dipolar modes propagating along a chain of metal nanoparticles are calculated by solving the full Maxwell equations, including radiation damping. The nanoparticles are treated as point dipoles, which means the results are valid only for $a/d \leq $ ⅓, where $a$ is the particle radius and $d$ the spacing. The discrete modes for a finite chain are first calculated, then these are mapped onto the dispersion relations appropriate for the infinite chain. Computed results are given for a chain of 50-nm diameter Ag spheres spaced by 75 nm. We find large deviations from previous quasistatic results: Transverse modes interact strongly with the light line. Longitudinal modes develop a bandwidth more than twice as large, resulting in a group velocity that is more than doubled. All modes for which $k_{mode} \leq \omega/c$ show strongly enhanced decay due to radiation damping.

# I. Introduction

The possibility of using metal nanoparticle chains to propagate optical excitations is attractive for integrated optics applications, since it can lead to optical processing circuitry with dimensions comparable to the wavelength of the light. Such small optical circuits are not possible with conventional integrated optics techniques, which are generally diffraction-limited in their size scale. The building blocks for chain waveguides are closely-spaced metal spheres or spheroids with sizes in the tens of nanometer range, which is an order of magnitude smaller than optical wavelengths. The fundamental excitations that lead to propagation are the dipolar resonances of the individual particles (also called Mie resonances or plasma resonances).[1] In practice these resonances are well-defined excitations only for noble metal spheroids and only in the frequency range for which the metallic behavior is free-electron-like, i.e., Re($\varepsilon$) < 0 and Im($\varepsilon$) < -Re($\varepsilon$), where $\varepsilon(\omega)$ is the complex metal dielectric response function.

Several recent studies investigate mode propagation in metal nanoparticle chains.[2,3,4,5,6] Quinten et al.[2] give numerical results for a chain of Ag spheres and find a 900 nm 1/$e$ intensity decay length for 50-nm diameter spheres, spaced by $d$ = 75 nm and excited at the dipole resonant frequency. Propagation is found only for longitudinal excitation, i.e., the dipole moments aligned along the chain direction. Perpendicular propagation is much more highly attenuated. Brongersma et al.[3] model a similar chain of Ag spheres, but they treat them as point dipoles and include only the lowest order quasistatic 1/$r^3$ interaction between the dipoles. They derive dispersion relations for modes in an infinite chain and find similar results for longitudinal and transverse propagation. Park and Stroud[4] allow for finite-sized metal spheres, but still within the quasistatic approximation, by including the higher order multipole fields in the interparticle interactions. These interactions become important when the particles are nearly touching. However, as long as $a \leq d/3$, where $a$ is the particle radius and $d$ the separation, they show that the point-dipole results are adequate for the lowest (dipolar) bands. Maier et al.[6] present results from finite-difference time-domain simulations for chains of 25-nm radius Au spheres spaced by 75 nm, which they claim yield dispersion relations correlating well with the point-dipole model.

In this paper we extend the results discussed above to include the full, time-dependent fields of the oscillating dipoles. We retain the point dipole approximation, which is certainly adequate for frequencies near the dipolar resonance and so long as the sphere separation is more than about three times the radius. However, the interactions between dipoles now contain terms varying as $1/r$ and $1/r^2$, in addition to the quasistatic $1/r^3$ term. We also include in the dipole polarizability the effects of radiation damping. These refinements to the theory lead to major changes in the nature of the modes. Modes near the Brillouin zone center, which are at the highest frequency for transverse propagation and the lowest for longitudinal, are now highly damped via radiation damping, even in the absence of loss in the metal. The modes for transverse propagation are drastically modified when their dispersion relation crosses the light line. This is the phase matching condition between the free photon and the dipolar chain mode, and it is not surprising that it leads to a strong effect. In Section II we consider an infinite chain and recover previous quasistatic results.[3,4,6] The infinite chain can be solved exactly in the quasistatic case, but the inclusion of retardation requires a difficult analytic continuation into the lower half frequency plane. We have found it simpler to avoid this difficulty by considering a finite chain, which we do in Section III. A chain of $N$ particles will have $N$ discrete normal modes. In the limit of large $N$, the normal mode frequencies form a continuous distribution, corresponding to the dispersion relation $\omega = \omega(k)$. We show how to find these modes and to map them onto the dispersion curve for an infinite chain. In Section IV we consider propagation of a driven excitation along the chain. We give some numerical results for various potential experiments and previous model calculations and finally provide a brief summary.

## II. Infinite chain

We begin with the standard expression for the electric field generated by a point dipole **p** oscillating with frequency $\omega$ ($\propto e^{-i\omega t}$)[7]

$$\mathbf{E}(\mathbf{r},t) = \left[ (1 - \frac{i\omega r}{c}) \frac{3\hat{\mathbf{r}} \cdot \mathbf{p}\hat{\mathbf{r}} - \mathbf{p}}{r^3} + \frac{\omega^2}{c^2} \frac{\mathbf{p} - \hat{\mathbf{r}} \cdot \mathbf{p}\hat{\mathbf{r}}}{r} \right] e^{i\omega r/c}. \tag{1}$$

Here **r** is the position vector pointing from the dipole to the field point. We should emphasize that this expression contains the full effects of retardation. For a linear chain of point dipoles spaced a distance $d$, in the absence of an applied field, the field at each dipole is the sum of the fields due to all the other dipoles. The induced moment on the $n^{\text{th}}$ dipole is the polarizability $\alpha(\omega)$ times this field,

$$\mathbf{p}_n = \alpha(\omega) \sum_{m \neq n} \left( (1 - \frac{i\omega|n-m|d}{c}) \frac{3\hat{\mathbf{r}} \cdot \mathbf{p}_m \hat{\mathbf{r}} - \mathbf{p}_m}{|n-m|^3 d^3} + \frac{\omega^2}{c^2} \frac{\mathbf{p}_m - \hat{\mathbf{r}} \cdot \mathbf{p}_m \hat{\mathbf{r}}}{|n-m|d} \right) e^{i\omega|n-m|d/c}. \tag{2}$$

Here we should keep in mind that $\hat{\mathbf{r}}$ points along the chain. We seek normal modes of the form of traveling waves $\mathbf{p}_n \propto e^{inkd}$. We then get a pair of dispersion relations, one for the transverse modes ($\mathbf{p}_n \perp \hat{\mathbf{r}}$) and one for the longitudinal modes ($\mathbf{p}_n \parallel \hat{\mathbf{r}}$),

$$1 + 2\frac{\alpha(\omega)}{d^3}\sum_{j=1}^{\infty}\left((1-i\frac{\omega d}{c}j)\frac{1}{j^3} - \frac{\omega^2 d^2}{c^2}\frac{1}{j}\right)\cos jkd\, e^{i\omega dj/c} = 0, \quad \text{(Transverse)} \tag{3a}$$

$$1 - 4\frac{\alpha(\omega)}{d^3}\sum_{j=1}^{\infty}\left((1-i\frac{\omega d}{c}j)\frac{1}{j^3}\right)\cos jkd\, e^{i\omega dj/c} = 0. \quad \text{(Longitudinal)} \tag{3b}$$

For $k$ real these equations are to be solved for the complex normal mode frequencies $\omega = \omega(k)$. Since the normal modes must be decaying in time, it is clear that these normal mode frequencies must have a negative imaginary part, $\text{Im}(\omega) \leq 0$. However, because of the factor $e^{i\omega dj/c}$, the sums converge only for $\text{Im}(\omega) \geq 0$. There is a prescription for dealing with this difficulty: evaluate the sums in the upper half-plane and then analytically continue them into the lower half-plane. In the following Section we show how to avoid this procedure by considering a finite chain. This difficulty disappears in the quasistatic approximation, which corresponds to forming the limit $c \to \infty$. Before discussing this limit, we must say something about the polarizability.

For a dielectric sphere in vacuum the quasistatic dipole polarizability has the form

$$\alpha(\omega) = \frac{\varepsilon(\omega)-1}{\varepsilon(\omega)+2}a^3, \tag{4}$$

where $a$ is the sphere radius and $\varepsilon(\omega)$ is the dielectric constant. For a metal sphere we use the Drude model for the dielectric response,[8] that is

$$\varepsilon(\omega) = 1 - \frac{\omega_P^2}{\omega(\omega + i\nu)}, \qquad (5)$$

where $\omega_p$ is the plasma frequency and $\nu$ the electron scattering rate. To be consistent with our inclusion of retardation in the dipole fields, we must include the effect of radiation reaction in the polarizability. This effect can be introduced through the usual prescription:[9]

$$\frac{1}{\alpha} \rightarrow \frac{1}{\alpha} - i\frac{2\omega^3}{3c^3}. \qquad (6)$$

Although this form can lead to acausal behavior in the response, it is satisfactory so long as the added term is small.[10]

With this form of the polarizability the dispersion relations (3) can be written in the form

$$\frac{\omega^2}{\omega_0^2}(1 + i\frac{2\omega\omega_0^2}{3c^3}a^3) + i\frac{\nu\omega}{\omega_0^2} = 1 + 2\frac{a^3}{d^3}\sum_{j=1}^{\infty}\left(1 - i\frac{\omega d j}{c} - \frac{\omega^2 d^2 j^2}{c^2}\right)\frac{\cos jkd}{j^3}e^{i\omega dj/c}, \qquad (7a)$$

$$\frac{\omega^2}{\omega_0^2}(1 + i\frac{2\omega\omega_0^2}{3c^3}a^3) + i\frac{\nu\omega}{\omega_0^2} = 1 - 4\frac{a^3}{d^3}\sum_{j=1}^{\infty}\left(1 - i\frac{\omega d j}{c}\right)\frac{\cos jkd}{j^3}e^{i\omega dj/c}, \qquad (7b)$$

where $\omega_0 = \omega_p/\sqrt{3}$ is the plasma resonance frequency for the sphere. Here, as in the equations (3), the upper equation is for the transverse case and the lower for the longitudinal. One can see by inspection that for real $kd$ these expressions are analytic in the upper half $\omega$-plane. The difficulty, as noted above, is with the evaluation in the lower half-plane, where the sums diverge.

We return to this difficulty in the following Section, but first we consider the quasistatic response for a lossless metal sphere, which corresponds to setting $v = 0$ and $c = \infty$. The polarizability (4) then takes on the simple form:

$$\frac{1}{\alpha(\omega)} = \frac{1}{a^3}(1 - \frac{\omega^2}{\omega_0^2}). \tag{8}$$

We can rewrite the dispersion relations (7) as

$$\frac{\omega^2}{\omega_0^2} = 1 + 2\frac{a^3}{d^3}\sum_{j=1}^{\infty}\frac{\cos jkd}{j^3}, \quad \text{(Transverse)} \tag{9a}$$

$$\frac{\omega^2}{\omega_0^2} = 1 - 4\frac{a^3}{d^3}\sum_{j=1}^{\infty}\frac{\cos jkd}{j^3}. \quad \text{(Longitudinal)} \tag{9b}$$

The sums in these equations can be easily evaluated to yield the dispersion relations plotted as solid curves in Fig. 1. These results are the same as those obtained earlier by Brongersma et al.[3] and Park and Stroud.[4]

## III. Finite chain

We remarked above that the infinite sums in the dispersion relations (3) do not converge for $\text{Im}(\omega) < 0$, where normal mode frequencies must lie. A simple way to avoid this problem is to consider a finite chain. As we show explicitly in the following, a chain of 20 spheres is adequate to obtain a dispersion curve. For a chain of $N$ spheres, equation (2) becomes a set of $N$ coupled equations in the $N$ unknown moments of the spheres. We write these equations in matrix form:

$$\mathbf{Mp} = 0, \tag{10}$$

where **p** is the *N*-rowed column vector of the dipole moments and the matrix **M** is defined by

$$\mathbf{M}_{n,n} = \frac{a^3}{\alpha(\omega)}, \quad n = 1, \cdots, N$$

$$\mathbf{M}_{n,m \neq n} = \frac{a^3}{d^3}\left(1 - i\frac{\omega|n-m|d}{c} - \frac{\omega^2|n-m|^2 d^2}{c^2}\right)\frac{e^{i\omega|n-m|d/c}}{|n-m|^3}, \quad \text{(transverse)} \quad (11)$$

$$\mathbf{M}_{n,m \neq n} = -2\frac{a^3}{d^3}\left(1 - i\frac{\omega|n-m|d}{c}\right)\frac{e^{i\omega|n-m|d/c}}{|n-m|^3}. \quad \text{(longitudinal)}$$

The normal modes correspond to the complex zeros of the determinant of **M**,

$$\det\{\mathbf{M}(\omega)\} = 0. \tag{12}$$

For a chain of *N* spheres there will be *N* normal modes. The problem is analogous to that of a chain of coupled oscillators in which each oscillator is coupled to all the others. We have used MATLAB® 6 to solve this problem for a chain of 20 spheres. For the purpose of comparison with previous results, we choose $a = 25$ nm and $d = 75$ nm. Since we are interested in the response near the dipolar resonance of an Ag sphere, we fix $\omega_p$ and $\nu$ to yield the optical constants of Ag at the resonance frequency $\hbar\omega_0 = 3.5$ eV, which gives $\hbar\omega_p = 6.18$ eV and $\hbar\nu = 0.7$ eV.[11] The results are shown in Tables 1 & 2, where we give the real and imaginary parts of the dimensionless normal mode frequencies $\omega/\omega_0$. The electron scattering loss is underestimated for these small particles, since we use bulk optical properties. However, this loss mechanism is still the primary cause of the mode damping.

The first column in the tables is the mode number, which we define as one plus the number of sign changes in the normal mode solution. To find this solution and, hence, the mode number, we solve the driven problem,

$$\mathbf{Mp} = \mathbf{v}, \tag{13}$$

with a simple choice for the column vector **v** (for example all but the first row equal to zero). Evaluating **M** at the resonance frequency, the normal mode solution is

$$\mathbf{p} = \mathbf{M}^{-1}\mathbf{v}. \tag{14}$$

Strictly speaking, if $\omega$ is exactly at the normal mode frequency, this solution will be infinite, since $\det(\mathbf{M}) = 0$, but in practice because of the small numerical inaccuracy the solution will be large but finite. In Fig. 2 we show the real part of this normal mode solution plotted versus sphere number for three examples, all for the transverse case. Corresponding plots for the longitudinal modes look identical. The top and bottom examples correspond, respectively, to the maximum (19) and minimum (0) number of sign changes. The middle curve shows one sign change. Clearly these mode solutions are suggestive of standing waves, and the mode number is easily determined. In the second column of the Tables we show a value of *kd* for each mode. We make this assignment with the formula

$$kd = \frac{(N-2)n + 1}{N(N-1)}\pi, \tag{15}$$

where n is the mode number. Note that for $n = 1$ this formula gives a wavelength $\lambda = 2Nd$, in agreement with the profile shown in the bottom of Fig. 2. The point in introducing this quantity is that we can view these calculated points as a discrete approximation to the continuous dispersion curve in the Brillouin zone for an infinite periodic array. Finally, in the remaining columns of the Tables, we give the real and imaginary parts of the calculated dimensionless normal mode frequencies, first with the inclusion of loss in the metal and then for the purely radiative case.

From the Tables we see that in the absence of loss in the metal, where all the loss is due to the radiated power, the points nearer the zone center have much larger imaginary parts than those near the zone boundary. This effect is much reduced for a lossy metal, where the imaginary parts are always large. The mode number dependence is a radiation effect that will be discussed more fully below.

In Fig.3 we plot results for the transverse case. The solid curve is the same quasistatic result shown in Fig.1, and the solid round points are the quasistatic normal mode frequencies obtained by the above method for a finite chain with 20 particles. The excellent agreement indicates that a finite chain of 20 particles is sufficient to reproduce the dispersion relation for an infinite chain. We expect this to be true even with the inclusion of the full retarded fields. The square points are a plot of the real values given in Table 1. The points are joined by dotted straight line segments to guide the eye. The nearly vertical dashed line is the light line $\omega = ck$. There is a dramatic deviation from the quasistatic result when the light line intersects the dispersion curve. At this point the dipolar modes are phase-matched to the free photon propagating along the chain at the same frequency. Finally, the triangular points correspond to an ideal metal ($v = 0$) with the same plasma frequency, showing the small effect on the dispersion curve of loss in the metal. However, metallic loss has a larger effect on the imaginary part of the mode frequency, especially for $kd$ near $\pi$.

In Fig. 4 we show the same results for the longitudinal case. The biggest difference here, compared with transverse excitation, is that there is no sharp interaction with the light line, since only transverse photons can propagate along the chain. There are also significant differences compared with the quasistatic approximation, e.g., the band width is nearly doubled by the inclusion of the full retarded fields and the group velocity near the band center is increased by more than a factor of two.

The effects of radiation damping are shown in Fig. 5, where we plot the imaginary part of the normal mode frequency as a function of $kd$ for a lossless metal chain. For any $k$-value such that $k_{mode} < \omega/c$, the array will generate strong far-field radiation at an angle $\theta$ to the chain axis where $\cos(\theta) = k_{mode} c/\omega$. This condition on $k$ occurs approximately at mode number 9, indicated by the dashed vertical line in Fig. 5 for the parameters we are using; it leads to a large increase in $-\text{Im}(\omega/\omega_0)$ for all lower modes. The result applies to both longitudinal and transverse modes.

## IV Propagation along a finite chain

In the previous Section we considered the problem of determining the complex normal mode frequencies. In terms of a dispersion relation $\omega = \omega(k)$, we there found the complex $\omega$ for real $k$. In discussing propagation we must, so to speak, invert this problem and study propagation in a chain of spheres driven with a real frequency. Consider, therefore, a chain in which the first sphere is driven with an applied optical field at frequency $\omega = \omega_0$. The column-vector of the dipole moments will then be given by Eq. (14), but now with the matrix **M** evaluated at the real driving frequency $\omega_0$ and with **v** the column vector in which all but the first row is zero. The result for a chain of 50 Ag spheres is shown in Fig. 6, in which the square of the absolute value of the dipole moment is plotted versus distance both for longitudinal and transverse excitations. Note that after an initial transient in which the decay is rapid and nonexponential, these log plots become approximately straight lines (with endpoint effects for the last few spheres in the chain). We fit a straight line to the points $n = 35$-$45$ and extract the decay lengths $\alpha^{-1}$ shown in the figure. Although the choice of points to use for the fit is somewhat arbitrary, this choice gives an excellent fit and it allows us to compare with the decay lengths obtained by other methods. To get the phase of the wave, we fit the calculated complex dipole moment to the form

$$p(x) = A \exp(ikd \frac{x}{d} - \frac{1}{2} \alpha x). \tag{16}$$

The results are shown in Fig. 7, where the real part of the dipole moment is plotted versus distance, again for longitudinal and transverse excitation. The fitted curves are shown as solid lines, the points are the solution of Eq. (14).

Quinten et al.[2] considered this same problem of propagation down a chain of 50 Ag spheres. They included the effect of higher multipoles, but restricted the calculation to the near field, which we interpret to mean the quasistatic approximation. Thus their calculation should be based on the same model as that of Park and Stroud.[4] But as shown by these latter authors, the effects of higher multipoles are negligible for the parameters chosen ($a/d = ⅓$). Therefore, we conclude that the differences between our results and those of Quinten et al. are entirely due to our inclusion of the effects of retardation. The most important difference is that Quinten et al. conclude that no significant propagation occurs for transverse excitation. On the contrary, as shown in Fig. 6, we find that although for transverse excitation the initial decay is more rapid, at long distances the decay is even slower than that for longitudinal excitation. Surprisingly, the 1/$e$ decay length they find for longitudinal excitation ($\alpha^{-1}$ = 900 nm) is comparable to ours (~ 700 nm).

Our quasistatic solution to the same chain of 50 Ag spheres shows a much faster initial decay. We also find that the initial decay rate is lowest when the chain is excited near the band center ($\omega = \omega_0$), but the rate far down the chain is rather insensitive to the driving frequency. However, this result depends on our use of the Drude model for the metal response, which is clearly not valid far from the band center.

Maier et al.[6] recently simulated propagation in a chain of 50-nm diameter Au spheres spaced by 75 nm using a finite-difference time-domain method that solves the full set of Maxwell's equations. They chose Drude parameters ($\hbar\omega_p$ = 4.47 eV, $\hbar\nu$ = 0.164 eV) to model the optical response of Au in the vicinity of the plasma resonance. However, these parameters do not fit the Au optical data of Johnson and Christy,[11] nor are they consistent with the simulated results shown in their Fig. 2.[12] These authors state that their simulated data "are in excellent agreement with the predictions from the point-dipole model". Since the model they refer to is quasistatic, this statement is difficult to reconcile with the fact that we find rather large differences between the full solution to Maxwell's equations and the quasistatic approximation.

A drawback of using spherical noble-metal particles is that the *d*-band absorption at the plasma resonance frequency significantly increases the loss in the metal above its free-electron value. This is especially true for Au and Cu and to a lesser extent for Ag. The solution to this problem, as discussed extensively in the literature on surface-enhanced Raman scattering,[13,14] is to use non-spherical particles. All of the methods developed above for spherical particle chains can be easily adapted to spheroidal particles with a single change. The expression for $\alpha(\omega)/a^3$ in Eq.(11) for the diagonal elements of the matrix **M** must be replaced by the general formula for a spheroid:

$$\frac{\alpha(\omega)}{a^3} \rightarrow \frac{1}{3}\frac{\varepsilon(\omega)-1}{1+[\varepsilon(\omega)-1][L-i\omega^3 V/6\pi c^3]}, \qquad (17)$$

where $V$ is the volume of the spheroid and $L$ is the depolarization factor.[1] For a sphere, $L = \frac{1}{3}$; for a prolate spheroid in the long direction or an oblate spheroid in the wide direction, $L$ will be less than $\frac{1}{3}$, and the plasma resonance will be shifted to lower frequencies. This frequency shift lowers both the *d*-band absorption loss and the radiative damping loss.

In summary, we have found the dispersion relations for dipolar modes propagating in a chain of metal nanoparticles. We use the point-dipole model for the fields, which means the results are valid when the sphere spacing is greater than or equal to about three times the sphere radius, and we solve the full Maxwell equations including the retarded fields. The effects of these retarded fields are quite striking compared with results from previous quasistatic treatments. In a lossless metal radiation damping affects all modes for which $k_{mode} < \omega/c$. Transverse modes are strongly perturbed when $k_{mode} \cong \omega/c$. Longitudinal modes develop a larger bandwidth, and their group velocity is more than twice its value in the quasistatic case.

Table 1. **Normal mode frequencies for transverse excitation in a chain of 20 Ag spheres of radius 25 nm spaced by 75 nm. The Ag optical response is modeled as $\varepsilon(\omega) = 1 - \omega_p^2/(\omega^2 + i\omega\nu)$ with $\hbar\omega_p = 6.18$ eV. ($\omega_0 = \omega_p/\sqrt{3}$)**

| Mode # | $kd$ | $\hbar\nu = 0.7$ eV | | $\hbar\nu = 0$ | |
|---|---|---|---|---|---|
| | | Re($\omega/\omega_0$) | Im($\omega/\omega_0$) | Re($\omega/\omega_0$) | Im($\omega/\omega_0$) |
| 1 | 0.157080 | 1.046665 | -0.168112 | 1.064838 | -0.062402 |
| 2 | 0.305892 | 1.044235 | -0.170080 | 1.062334 | -0.064203 |
| 3 | 0.454704 | 1.039432 | -0.173101 | 1.057937 | -0.067202 |
| 4 | 0.603516 | 1.033535 | -0.178439 | 1.051393 | -0.071448 |
| 5 | 0.752329 | 1.022636 | -0.183155 | 1.041861 | -0.076846 |
| 6 | 0.901141 | 1.011361 | -0.203115 | 1.028288 | -0.083853 |
| 7 | 1.049953 | 0.988978 | -0.199264 | 1.006917 | -0.091521 |
| 8 | 1.198766 | 0.874471 | -0.134102 | 0.947637 | -0.087735 |
| 9 | 1.347578 | 0.920553 | -0.095956 | 0.946439 | -0.032017 |
| 10 | 1.496390 | 0.950655 | -0.085079 | 0.960098 | -0.011306 |
| 11 | 1.645202 | 0.966177 | -0.087880 | 0.970231 | -0.003948 |
| 12 | 1.794015 | 0.972249 | -0.091193 | 0.975668 | -0.001455 |
| 13 | 1.942827 | 0.974841 | -0.093493 | 0.978102 | -0.000002 |
| 14 | 2.091639 | 0.975637 | -0.095173 | 0.979818 | -0.000048 |
| 15 | 2.240452 | 0.975533 | -0.096463 | 0.979846 | -0.000003 |
| 16 | 2.389264 | 0.974997 | -0.097491 | 0.979184 | -0.000041 |
| 17 | 2.538076 | 0.974322 | -0.098302 | 0.978911 | -0.000077 |
| 18 | 2.686888 | 0.973670 | -0.098912 | 0.978102 | -0.000002 |
| 19 | 2.835701 | 0.973143 | -0.099339 | 0.977560 | -0.000044 |
| 20 | 2.984513 | 0.972804 | -0.099592 | 0.977291 | -0.000002 |

Table 2. **Normal mode frequencies for longitudinal excitation in a chain of 20 Ag spheres of radius 25 nm spaced by 75 nm. The Ag optical response is modeled as $\varepsilon(\omega) = 1-\omega_p^2/(\omega^2+i\omega\nu)$ with $\hbar\omega_p = 6.18$ eV. ($\omega_0 = \omega_p/\sqrt{3}$)**

| Mode # | $kd$ | $\hbar\nu = 0.7$ eV | | $\hbar\nu = 0$ | |
|---|---|---|---|---|---|
| | | Re($\omega/\omega_0$) | Im($\omega/\omega_0$) | Re($\omega/\omega_0$) | Im($\omega/\omega_0$) |
| 1 | 0.157080 | 0.842932 | -0.186396 | 0.870986 | -0.092440 |
| 2 | 0.305892 | 0.846043 | -0.181791 | 0.873955 | -0.088063 |
| 3 | 0.454704 | 0.851659 | -0.173807 | 0.879052 | -0.080872 |
| 4 | 0.603516 | 0.859481 | -0.163724 | 0.886494 | -0.071168 |
| 5 | 0.752329 | 0.870402 | -0.149855 | 0.896579 | -0.059159 |
| 6 | 0.901141 | 0.884943 | -0.135540 | 0.909791 | -0.045555 |
| 7 | 1.049953 | 0.902368 | -0.117555 | 0.926571 | -0.030938 |
| 8 | 1.198766 | 0.929751 | -0.099795 | 0.947985 | -0.017313 |
| 9 | 1.347578 | 0.962603 | -0.091531 | 0.972423 | -0.007940 |
| 10 | 1.496390 | 0.992058 | -0.092911 | 0.996191 | -0.003509 |
| 11 | 1.645202 | 1.013288 | -0.097271 | 1.016871 | -0.002008 |
| 12 | 1.794015 | 1.030228 | -0.099193 | 1.034535 | -0.001324 |
| 13 | 1.942827 | 1.045850 | -0.100711 | 1.049957 | -0.000885 |
| 14 | 2.091639 | 1.058989 | -0.102093 | 1.063351 | -0.000614 |
| 15 | 2.240452 | 1.070506 | -0.102999 | 1.074867 | -0.000409 |
| 16 | 2.389264 | 1.080129 | -0.103886 | 1.084567 | -0.000269 |
| 17 | 2.538076 | 1.088007 | -0.104473 | 1.092481 | -0.000162 |
| 18 | 2.686888 | 1.094132 | -0.104980 | 1.098630 | -0.000089 |
| 19 | 2.835701 | 1.098497 | -0.105302 | 1.103018 | -0.000038 |
| 20 | 2.984513 | 1.101121 | -0.105505 | 1.105651 | -0.000009 |

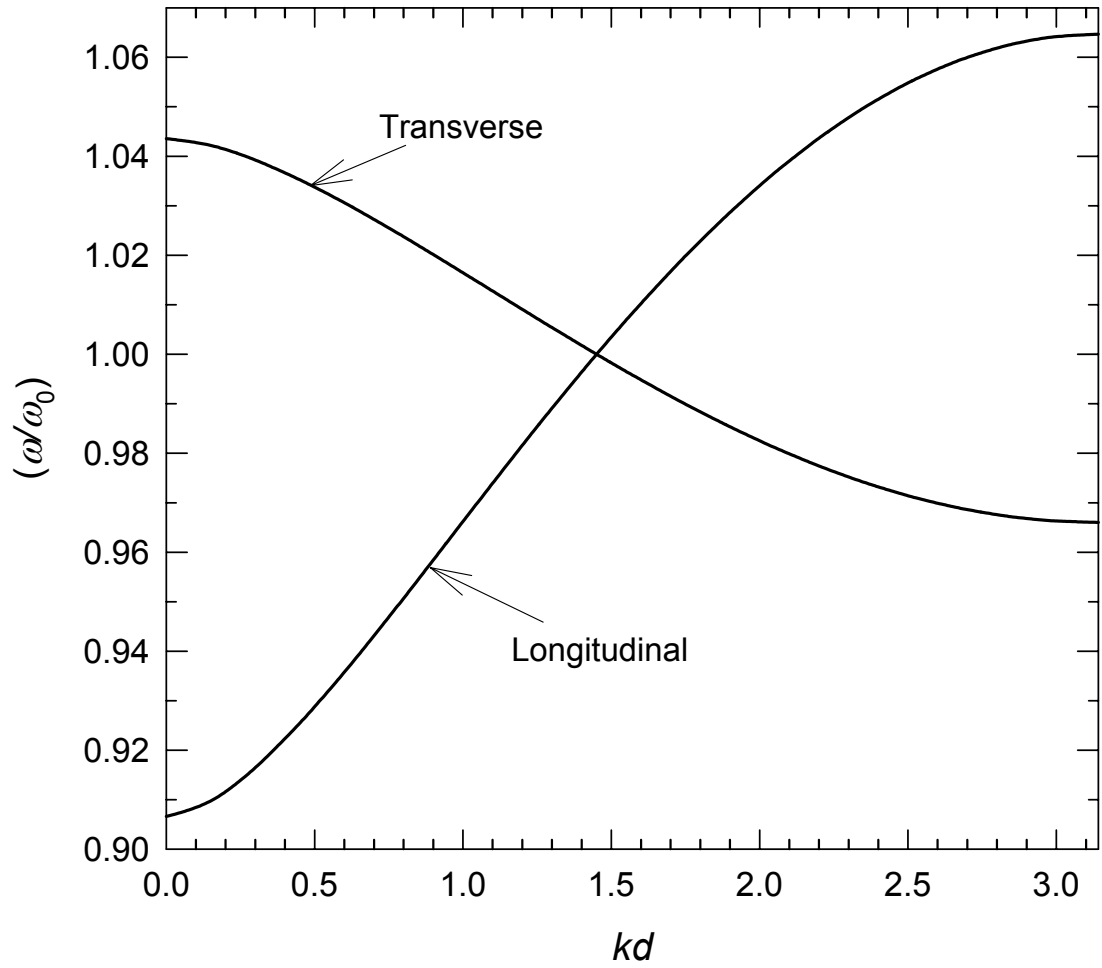

Fig. 1, Weber & Ford

Figure 1. Dispersion relations for dipolar modes in the quasistatic approximation for an infinite chain of 50-nm diameter Ag spheres spaced by 75 nm. A lossless Drude response is assumed with $\hbar\omega_p = 6.18$ eV.

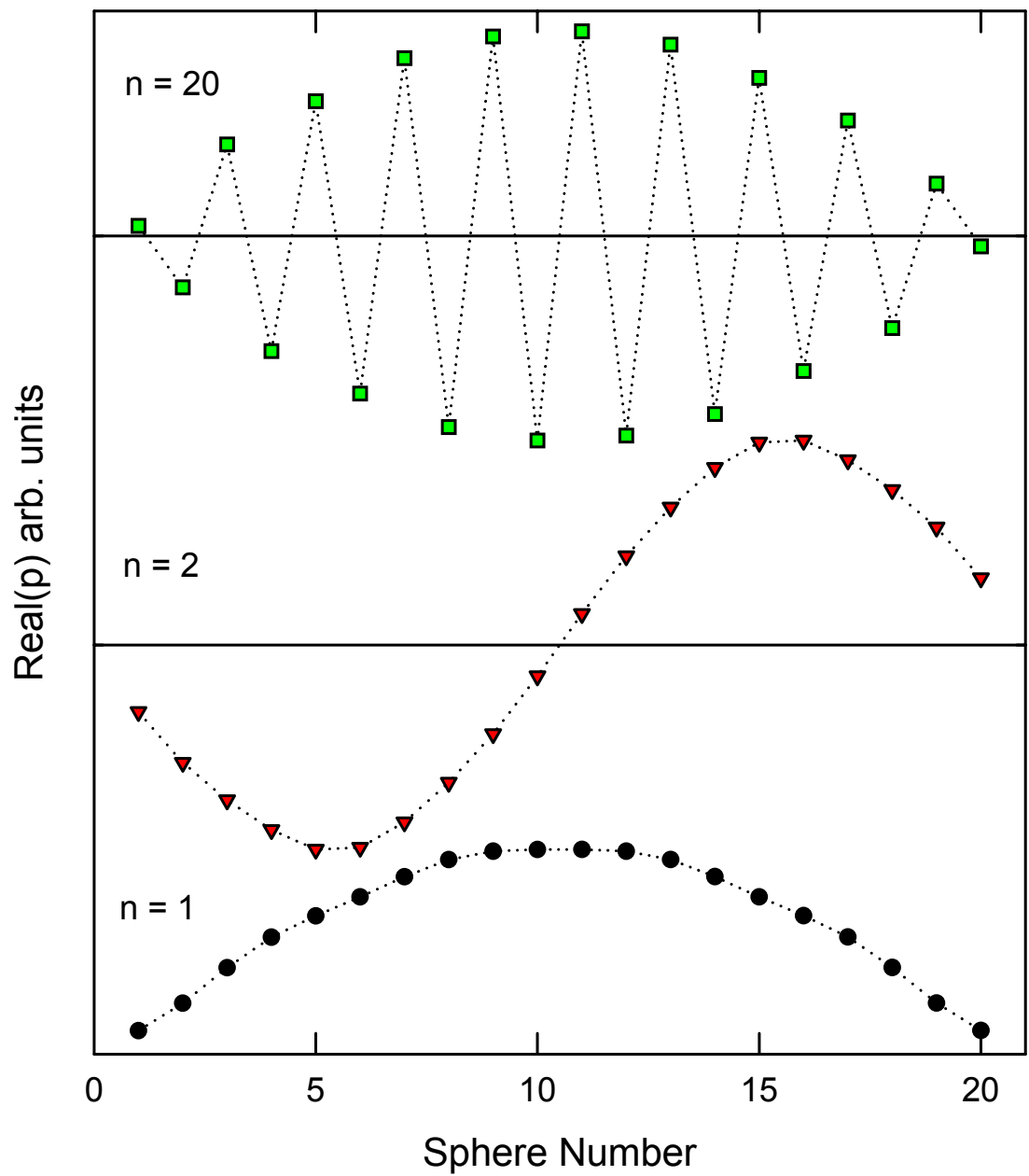

Figure 2. Mode profiles for transverse excitation in a chain of 20 50-nm diameter Ag spheres spaced by 75 nm. The horizontal line for each mode corresponds to Real(p) = 0.

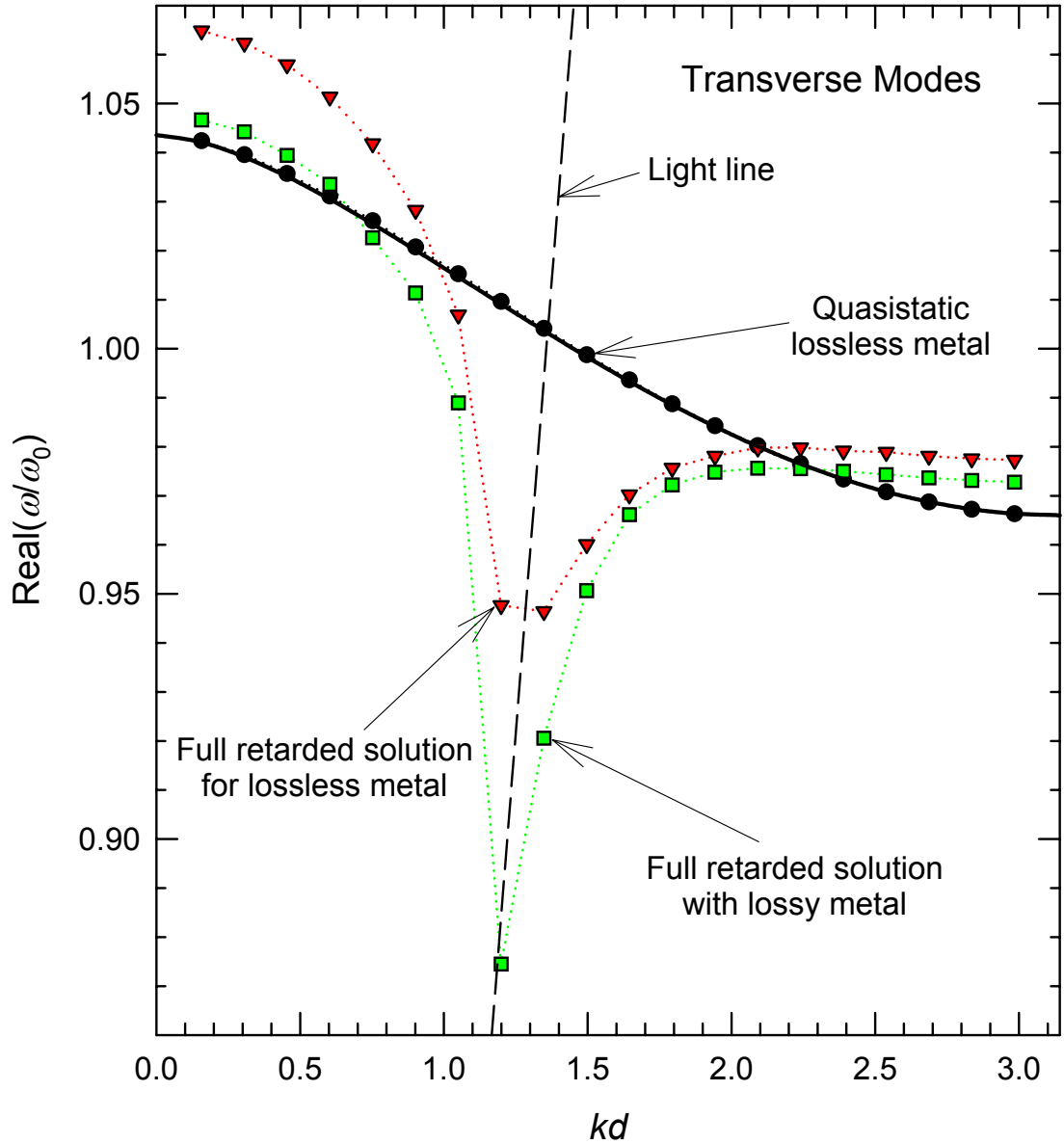

Fig. 3, Weber & Ford

Figure 3. Dispersion curves for transverse excitation as described in the text. Solid line is the same curve in Fig. 1 for an infinite chain. Points are for a finite 20-sphere chain: black circles (●) for the quasistatic approximation, green squares (■) for the full retarded solution with a lossy metal, red triangles (▼) for the full solution and an ideal metal. Dashed line is the light line, $\omega = ck$.

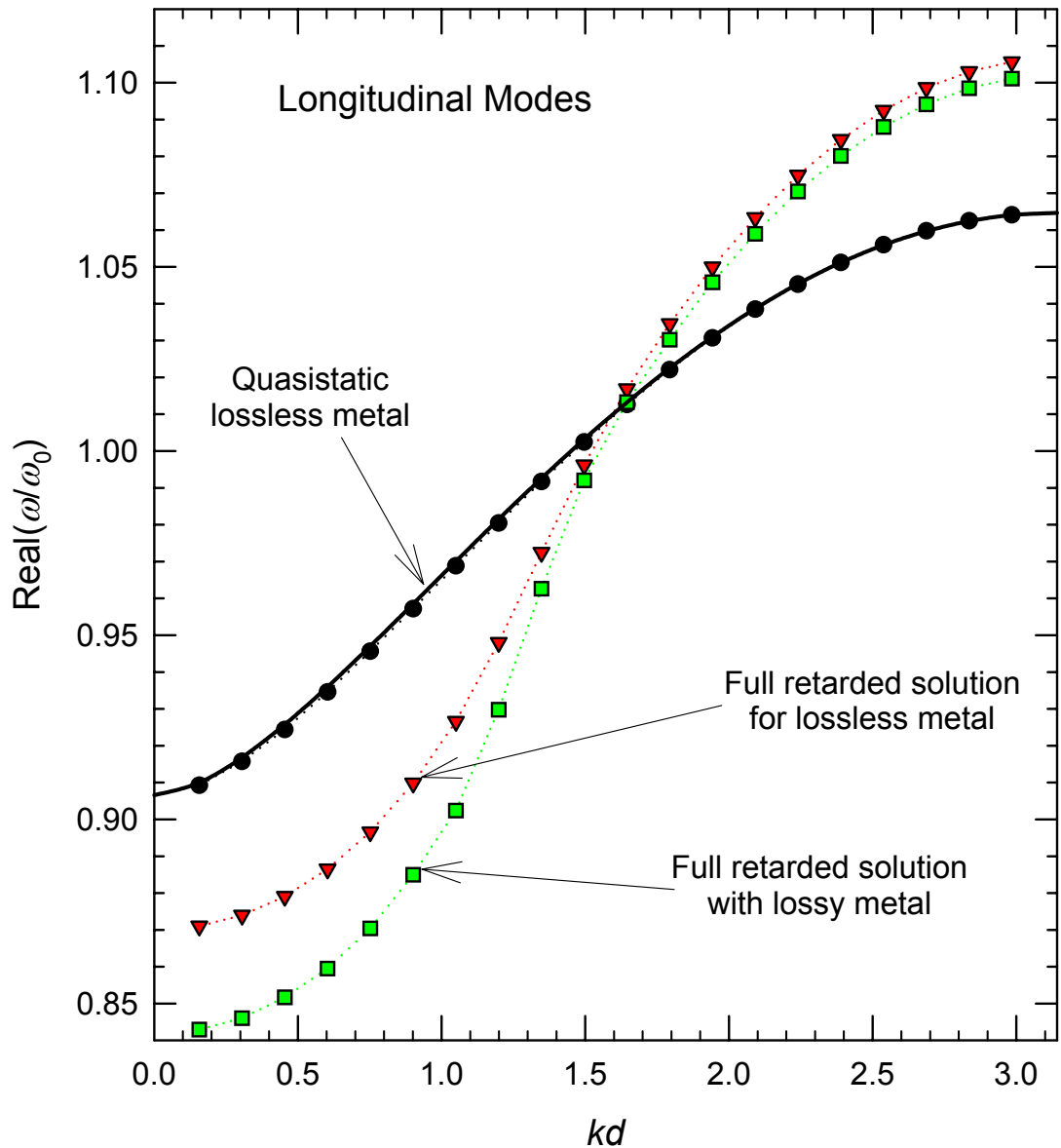

Figure 4. Same plots as in Fig. 3 for longitudinal excitation. Note the increased bandwidth associated with the full solution.

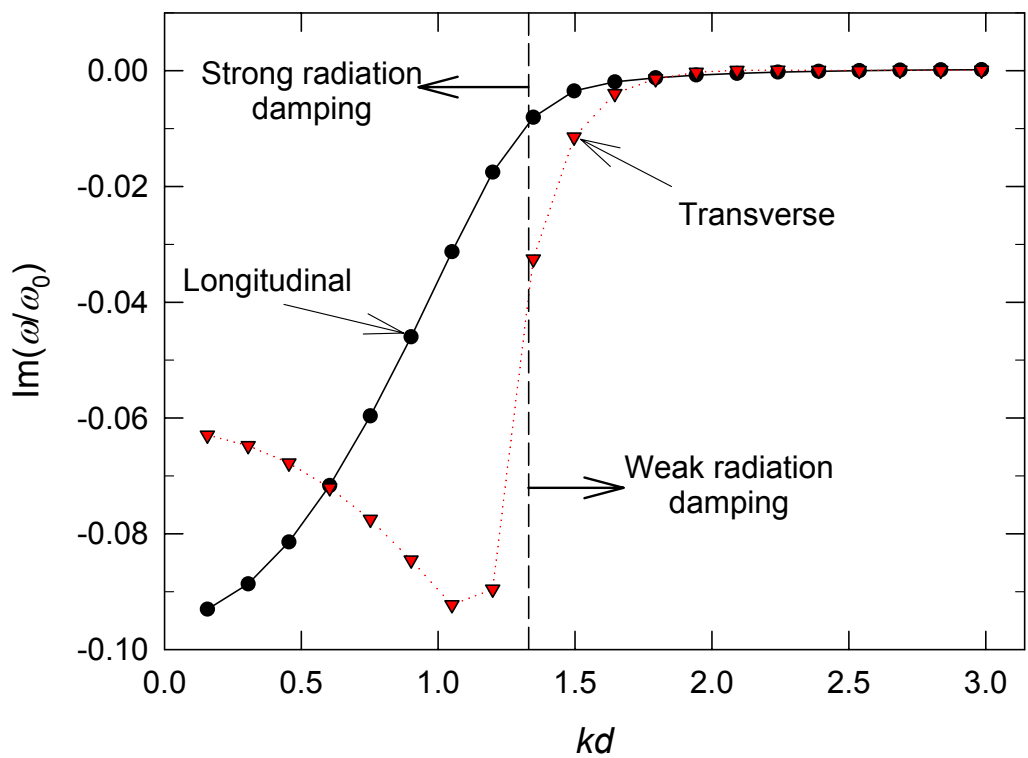

Fig. 5, Weber & Ford

Figure 5. Imaginary part of the normal mode frequencies from the last columns in Tables 1 (red triangles) and 2 (black circles) for a 20-sphere chain of lossless metal particles. The dashed vertical line corresponds approximately to $k_{mode} = \omega/c$.

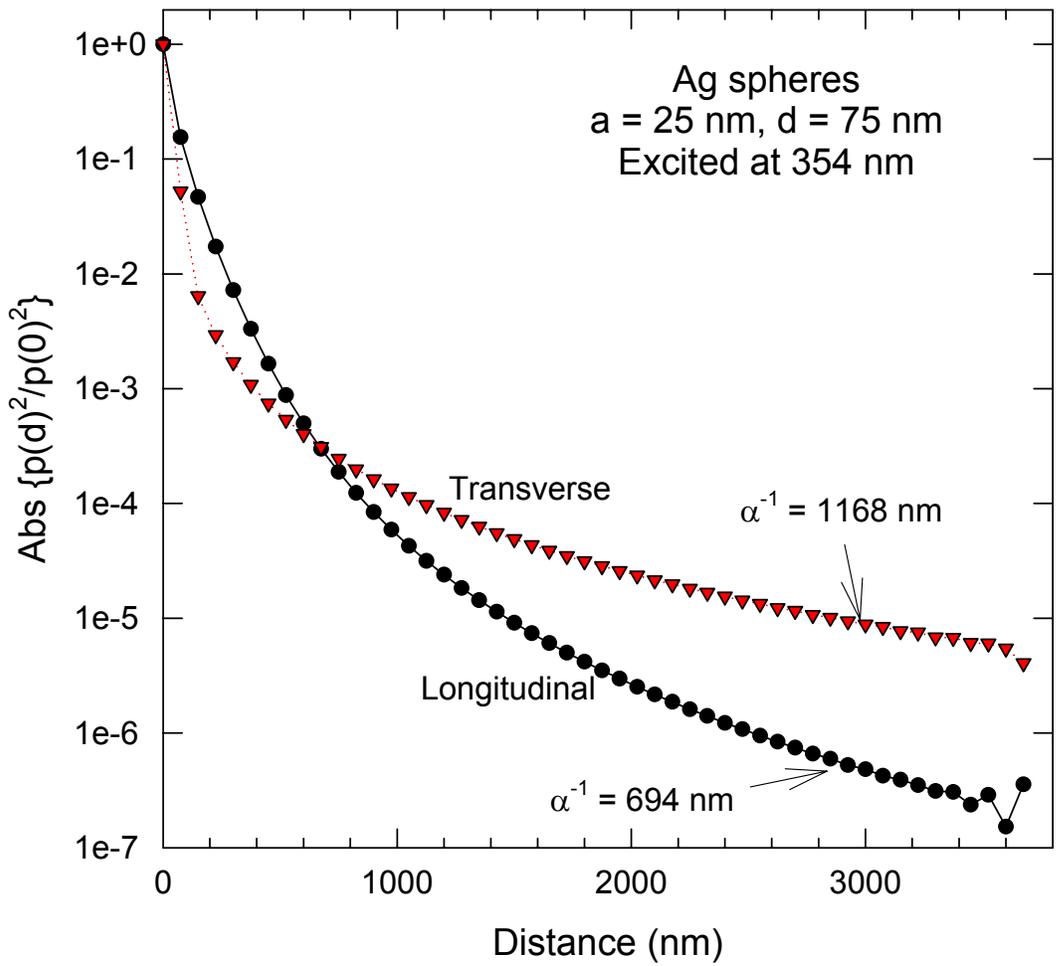

Figure 6. Normalized intensity (induced moment squared) when the first particle in a 50 particle chain is excited at the plasma resonance frequency. Attenuation coefficients determined from the slopes in the indicated regions.

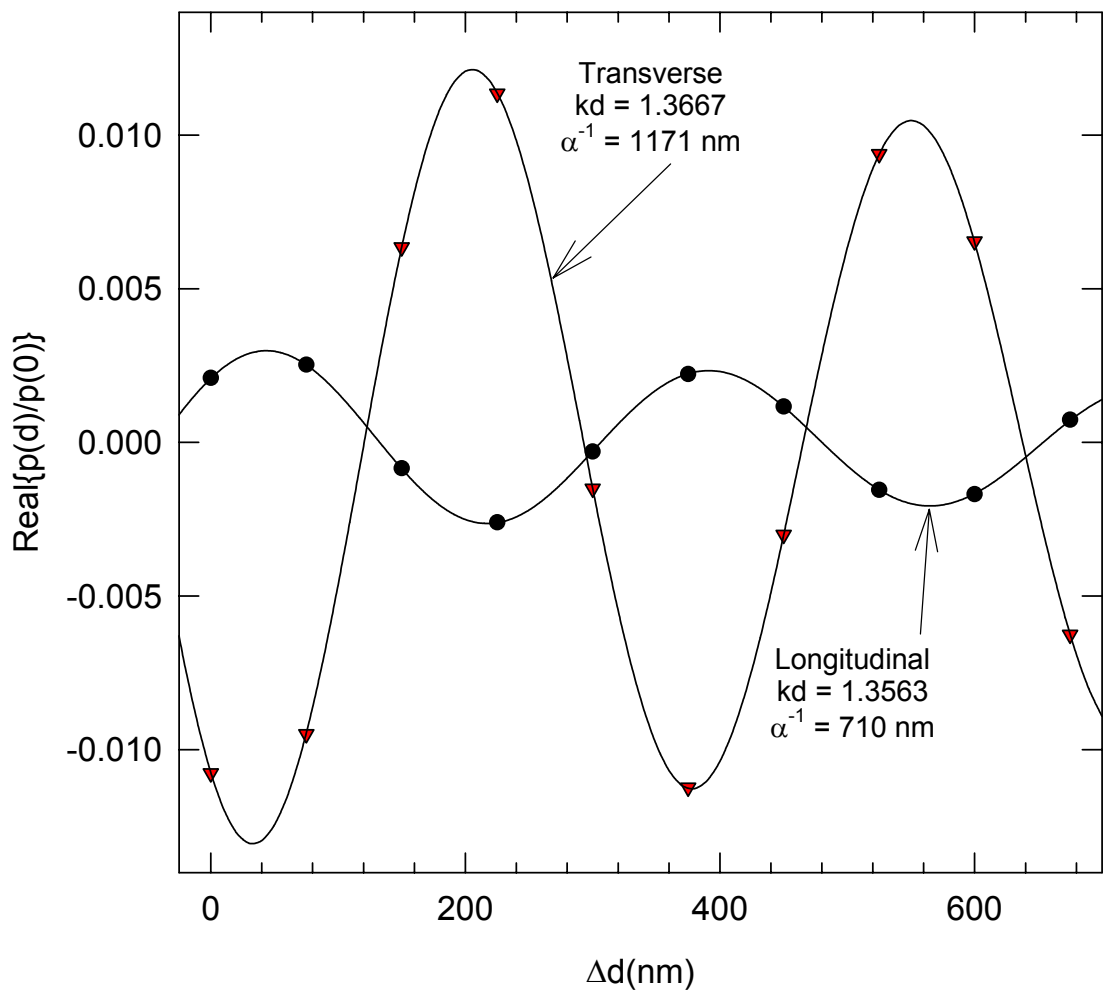

Figure 7. Points are calculated induced moments near the end of a 50-sphere chain as in Fig. 6; solid lines are fits to a decaying sine wave as described in the text.

the point at which Re($\varepsilon$) = -2 for Au from the ref. 11 optical data. However, the same Drude parameters yield Im($\varepsilon$) = 0.19, compared with Im($\varepsilon$) = 4.3 from ref.11. Moreover, Fig. 2 of ref. 6 shows $\hbar\omega_0$ = 2.40 eV instead of 2.59 eV.

[13] P. W. Barber, R. K. Chang, and H. Massoudi, Phys. Rev. B **27**, 7251 (1983).

[14] G. W. Ford and W. H. Weber, Physics Reports **113**, 195 (1984).